# *MotifbreakR v2*: extended capability and database integration

July 3, 2024

Simon G. Coetzee[1] & Dennis J. Hazelett[1,2]
1 Department of Computational Biomedicine at Cedars-Sinai Medical Center
2 Cancer Prevention and Control - Samuel Oschin Cancer Center, Cedars-Sinai

# Abstract

## Summary

*MotifbreakR* is a software tool that scans genetic variants against position weight matrices of transcription factors (TF) to determine the potential for the disruption of TF binding at the site of the variant. It leverages the Bioconductor suite of software packages and annotations to operate across a diverse array of genomes and motif databases. Initially developed to interrogate the effect of single nucleotide variants (common and rare SNVs) on potential TF binding sites, in *motifbreakR* v2, we have updated the functionality. New features include the ability to query other types of more complex genetic variants, such as short insertions and deletions (indels). This function allows modeling a more extensive array of variants that may have more significant effects on TF binding. Additionally, while TF binding is based partly on sequence preference, predictions of TF binding based on sequence preference alone can indicate many more potential binding events than observed. Adding information from DNA-binding sequencing datasets lends confidence to motif disruption prediction by demonstrating TF binding in cell lines and tissue types. Therefore, *motifbreakR* implements querying the ReMap2022 database for evidence that a TF matching the disrupted motif binds over the disrupting variant. Finally, in *motifbreakR*, in addition to the existing interface, we have implemented an R/Shiny graphical user interface to simplify and enhance access to researchers with different skill sets.

## Availability and implementation

*MotifbreakR* is implemented in R. Source code, documentation, and tutorials are available on Bioconductor at https://bioconductor.org/packages/release/bioc/html/motifbreakR.html and GitHub at https://github.com/Simon-Coetzee/motifBreakR

# 1 Introduction

Prediction of the likely consequences of variants on transcription factor (TF) binding is extremely valuable for hypothesis generation in human genetics and disease research and in the study of regulatory genomics broadly [1–5]. Tools and workflows have become increasingly sophisticated using machine learning and massively parallel reporter assays [6–9]. However, it remains practical to maintain software libraries that can generate predictions based on position weight matrix (PWM) based matching analysis as a first-pass hypothesis generator or part of a larger bioinformatics workflow. *MotifbreakR* remains relevant, having been used in many studies of human disease, from neurodegenerative disorders to COVID-19 research [9–59]. Additionally, it has been used in a variety of studies of model organisms, including yeast, mice, pigs, and blind cavefish [60–64]. It has even been used in paleoanthropology [65] and evolutionary genomics [66].

We originally published *motifbreakR* for variant analysis of genome-wide association studies (GWAS) [67]. What *motifbreakR* does well, which has led to its continued use across fields, is enable a standardized analysis of any genome curated by Bioconductor, using any available PWM library and any conceivable format for input variants, including BED format, VCFs, lists of rsIDs, and even "custom" variants specified in a flexible BED derived format. However, the original *motifbreakR* did not accept other types of more complex genetic variants, such as short insertions and deletions (indels) or variants with more than one alternative allele that constitute a significant fraction of variation: ~17% of variants found in whole-genome sequencing in gnomAD and ~23% of common variants (global MAF > 0.01) annotated in dbSNP. In addition, we found that incorporating published ChIP-seq data greatly enhances the impact of individual findings. Here, we added the ability to query indels, automated the complementary identification of cognate TF ChIP-seq peaks, and created a graphical user interface to make all these pipelines accessible to non-bioinformaticians.

# 2 Features

## 2.1 Insertion-Deletion Variants

We have implemented the import and analysis of indel variants to allow for querying a broader array of variations that alter transcription factor binding in the genome. Variants are scored by scanning a motif across the reference and the alternate sequence; the returned score is the highest-scoring match (or, equivalently, the match with the lowest p-value) in the whole sequence. The effect size is the difference between the best match on the reference allele and the alternate allele. The biggest challenge in implementation is defining a coherent coordinate system for specifying the position of the matching motif. In an example where an insertion is longer than the sequence of the queried PWM, the motif could be spliced and thus destroyed by the insertion. Alternatively, it creates the motif such that PWM overlaps the beginning, is entirely contained within, or overlaps the end of the insertion. In the instances where the motif is created, the position cannot be described by coordinates on the reference genome. *MotifbreakR* defines the coordinates of a motif relative to the edges of the indel. The first number is the number of bases upstream (negative) or downstream (positive) of the start of the variant describing where the motif starts. The second number is the number of bases upstream or downstream of the end of the variant, indicating where the motif ends (**Figure 1A**).

## 2.2 Querying of Transcription Factor Binding Database

While computational prediction of differential transcription factor binding potential based on sequence preference is the core of *motifbreakR*, grounding the analysis in observed transcription factor binding can improve the prioritization of results. To this end, we incorporate querying the ReMap2022 database of DNA-binding sequencing datasets to determine in-vitro evidence for transcription factor binding at the location of a *motifbreakR* result. In keeping with the existing flexibility of the *motifbreakR* package, with regards to species being investigated, one may query transcription factor binding data in *Homo sapiens* (hg38, or hg19 as liftover), *Mus musculus* (mm10, or mm39 as liftover), *Drosophila melanogaster* (dm6), and *Arabidopsis thaliana* (TAIR10).

The new functions empower the user to build a local database and query a motifbreakR result. Once built, *motifbreakR* can rapidly annotate its results with ReMap sourced transcription factor peaks corresponding to motif/transcription factor relationships provided by the constituent public *MotifDb* [68] sources. The user may optionally query an expanded motif/transcription factor relationship encompassing the entire potential transcription factor family as implemented by *MotifDb* based on *TFClass* [68,69]. Though not comprehensive, including the experiment biotype (cell lines and tissue types) from ReMap may also be helpful for hypothesis generation. Figure 1B shows an example of a variant rs143969848 that breaks a CTCF motif centered on a ChIP-seq peak in multiple cell lines.

## 2.3 Graphical User Interface

To make *motifbreakR* accessible to bench scientists, researchers with various skill sets, and others wishing to explore its capabilities on the web, we developed an R/Shiny-based graphical user interface (GUI) that facilitates all functions of the underlying *motifbreakR* package. In a workflow mirroring the code-based method, the user specifies individual rsIDs or "custom" variants and performs downstream tasks via pulldown menus and radio buttons. Like the R package version, users may upload VCFs or lists of SNVs in BED format. Any figures generated are downloadable/saved to the local environment. Finally, to promote reproducible science, analysis performed with the GUI will be output as code in R markdown format for publication.

## 2.4 Useful exports

*MotifbreakR* now exports results in various tabular formats compatible with database programs, including Excel or SQL. In addition, *motifbreakR* can export a BED file of user-defined subsets of top matches for display in browsers such as UCSC genome browser [70], WashU Epigenome browsers [71,72], or IGV [73,74]. Optionally, matches can be color-coded by motif quality (p-value-oriented) or disruptiveness (score-oriented).

# 3 Conclusion

We have updated *motifbreakR* to version 2 to include several new useful features, most notably indels, a new analysis pipeline to reference published ChIP-seq experiments that match *motifbreakR* predictions, and a GUI to promote accessibility and reproducibility.

# Declaration of competing interests

None to Declare.

# Acknowledgments

The authors would like to acknowledge funding from the following sources: NIH grants to Jason H. Moore (R01LM010098 & U01AG066833).

# References

1. Peña-Martínez EG, Rodríguez-Martínez JA. Decoding Non-coding Variants: Recent Approaches to Studying Their Role in Gene Regulation and Human Diseases. Front Biosci . 2024;16: 4.

2. Degtyareva AO, Antontseva EV, Merkulova TI. Regulatory SNPs: Altered Transcription Factor Binding Sites Implicated in Complex Traits and Diseases. Int J Mol Sci. 2021;22. doi:10.3390/ijms22126454

3. Coetzee GA. Understanding Non-Mendelian Genetic Risk. Curr Genomics. 2019;20: 322–324.

4. Hazelett DJ, Conti DV, Han Y, Al Olama AA, Easton D, Eeles RA, et al. Reducing GWAS Complexity. Cell Cycle. 2016;15: 22–24.

5. Purdue MP, Dutta D, Machiela MJ, Gorman BR, Winter T, Okuhara D, et al. Multi-ancestry genome-wide association study of kidney cancer identifies 63 susceptibility regions. Nat Genet. 2024;56: 809–818.

6. Han D, Li Y, Wang L, Liang X, Miao Y, Li W, et al. Comparative analysis of models in predicting the effects of SNPs on TF-DNA binding using large-scale in vitro and in vivo data. Brief Bioinform. 2024;25. doi:10.1093/bib/bbae110

7. Long E, Yin J, Funderburk KM, Xu M, Feng J, Kane A, et al. Massively parallel reporter assays and variant scoring identified functional variants and target genes for melanoma loci and highlighted cell-type specificity. Am J Hum Genet. 2022;109: 2210–2229.

8. Mulvey B, Dougherty JD. Transcriptional-regulatory convergence across functional MDD risk variants identified by massively parallel reporter assays. Transl Psychiatry. 2021;11: 403.

9. Choi J, Zhang T, Vu A, Ablain J, Makowski MM, Colli LM, et al. Massively parallel reporter assays of melanoma risk variants identify MX2 as a gene promoting melanoma. Nat Commun. 2020;11: 2718.

10. Booms A, Pierce SE, van der Schans EJC, Coetzee GA. Parkinson's disease risk enhancers in microglia. iScience. 2024;27: 108921.

11. McAfee JC, Lee S, Lee J, Bell JL, Krupa O, Davis J, et al. Systematic investigation of allelic regulatory activity of schizophrenia-associated common variants. Cell Genom. 2023;3: 100404.

12. Selewa A, Luo K, Wasney M, Smith L, Sun X, Tang C, et al. Single-cell genomics improves the discovery of risk variants and genes of atrial fibrillation. Nat Commun. 2023;14: 4999.

13. Benaglio P, Newsome J, Han JY, Chiou J, Aylward A, Corban S, et al. Mapping genetic effects on cell type-specific chromatin accessibility and annotating complex immune trait variants using single nucleus ATAC-seq in peripheral blood. PLoS Genet. 2023;19: e1010759.

14. Aygün N, Liang D, Crouse WL, Keele GR, Love MI, Stein JL. Inferring cell-type-specific causal gene regulatory networks during human neurogenesis. Genome Biol. 2023;24: 130.

15. Lutz MW, Chiba-Falek O. Bioinformatics pipeline to guide post-GWAS studies in Alzheimer's: A new catalogue of disease candidate short structural variants. Alzheimers Dement. 2023;19: 4094–4109.

16. Fabo T, Khavari P. Functional characterization of human genomic variation linked to polygenic diseases. Trends Genet. 2023;39: 462–490.

17. Benaglio P, Zhu H, Okino M-L, Yan J, Elgamal R, Nariai N, et al. Type 1 diabetes risk genes mediate pancreatic beta cell survival in response to proinflammatory cytokines. Cell Genom. 2022;2: 100214.

18. Zhang B, Zhang Z, Koeken VACM, Kumar S, Aillaud M, Tsay H-C, et al. Altered and allele-specific open chromatin landscape reveals epigenetic and genetic regulators of innate immunity in COVID-19. Cell Genom. 2023;3: 100232.

19. Prahl JD, Pierce SE, van der Schans EJC, Coetzee GA, Tyson T. The Parkinson's disease variant rs356182 regulates neuronal differentiation independently from alpha-synuclein. Hum Mol Genet. 2023;32: 1–14.

20. Ali MW, Chen J, Yan L, Wang X, Dai JY, Vaughan TL, et al. A risk variant for Barrett's esophagus and esophageal adenocarcinoma at chr8p23.1 affects enhancer activity and implicates multiple gene targets. Hum Mol Genet. 2022;31: 3975–3986.

21. Pahl MC, Le Coz C, Su C, Sharma P, Thomas RM, Pippin JA, et al. Implicating effector genes at COVID-19 GWAS loci using promoter-focused Capture-C in disease-relevant immune cell types. Genome Biol. 2022;23: 125.

22. Schilder BM, Raj T. Fine-mapping of Parkinson's disease susceptibility loci identifies putative causal variants. Hum Mol Genet. 2022;31: 888–900.

23. Wang T, Song J, Qu M, Gao X, Zhang W, Wang Z, et al. Integrative Epigenome Map of the Normal Human Prostate Provides Insights Into Prostate Cancer Predisposition. Front Cell Dev Biol. 2021;9: 723676.

24. Aygün N, Elwell AL, Liang D, Lafferty MJ, Cheek KE, Courtney KP, et al. Brain-trait-associated variants impact cell-type-specific gene regulation during neurogenesis. Am J Hum Genet. 2021;108: 1647–1668.

25. D'Antona S, Bertoli G, Castiglioni I, Cava C. Minor Allele Frequencies and Molecular Pathways Differences for SNPs Associated with Amyotrophic Lateral Sclerosis in Subjects Participating in the UKBB and 1000 Genomes Project. J Clin Med Res. 2021;10. doi:10.3390/jcm10153394

26. Xu M, Mehl L, Zhang T, Thakur R, Sowards H, Myers T, et al. A UVB-responsive common variant at chromosome band 7p21.1 confers tanning response and melanoma risk via regulation of the aryl hydrocarbon receptor, AHR. Am J Hum Genet. 2021;108: 1611–1630.

27. Pluta J, Pyle LC, Nead KT, Wilf R, Li M, Mitra N, et al. Identification of 22 susceptibility loci associated with testicular germ cell tumors. Nat Commun. 2021;12: 4487.

28. Liang D, Elwell AL, Aygün N, Krupa O, Wolter JM, Kyere FA, et al. Cell-type-specific effects of genetic variation on chromatin accessibility during human neuronal differentiation. Nat Neurosci. 2021;24: 941–953.

29. Novikova G, Kapoor M, Tcw J, Abud EM, Efthymiou AG, Chen SX, et al. Integration of Alzheimer's disease genetics and myeloid genomics identifies disease risk regulatory elements and genes. Nat Commun. 2021;12: 1610.

30. Su C, Argenziano M, Lu S, Pippin JA, Pahl MC, Leonard ME, et al. 3D promoter architecture re-organization during iPSC-derived neuronal cell differentiation implicates target genes for neurodevelopmental disorders. Prog Neurobiol. 2021;201: 102000.

31. Kristjánsdóttir K, Dziubek A, Kang HM, Kwak H. Population-scale study of eRNA transcription reveals bipartite functional enhancer architecture. Nat Commun. 2020;11: 5963.

32. Bao EL, Nandakumar SK, Liao X, Bick AG, Karjalainen J, Tabaka M, et al. Inherited myeloproliferative neoplasm risk affects haematopoietic stem cells. Nature. 2020;586: 769–775.

33. Jones MR, Peng P-C, Coetzee SG, Tyrer J, Reyes ALP, Corona RI, et al. Ovarian Cancer Risk Variants Are Enriched in Histotype-Specific Enhancers and Disrupt Transcription Factor Binding Sites. Am J Hum Genet. 2020;107: 622–635.

34. Vuckovic D, Bao EL, Akbari P, Lareau CA, Mousas A, Jiang T, et al. The Polygenic and Monogenic Basis of Blood Traits and Diseases. Cell. 2020;182: 1214–1231.e11.

35. Gagliardi A, Porter VL, Zong Z, Bowlby R, Titmuss E, Namirembe C, et al. Analysis of Ugandan cervical carcinomas identifies human papillomavirus clade-specific epigenome and transcriptome landscapes. Nat Genet. 2020;52: 800–810.

36. Corona RI, Seo J-H, Lin X, Hazelett DJ, Reddy J, Fonseca MAS, et al. Non-coding somatic mutations converge on the PAX8 pathway in ovarian cancer. Nat Commun. 2020;11: 2020.

37. Schulze KV, Swaminathan S, Howell S, Jajoo A, Lie NC, Brown O, et al. Edematous severe acute malnutrition is characterized by hypomethylation of DNA. Nat Commun. 2019;10: 5791.

38. Speedy HE, Beekman R, Chapaprieta V, Orlando G, Law PJ, Martín-García D, et al. Insight into genetic predisposition to chronic lymphocytic leukemia from integrative epigenomics. Nat Commun. 2019;10: 3615.

39. Booms A, Coetzee GA, Pierce SE. MCF-7 as a Model for Functional Analysis of Breast Cancer Risk Variants. Cancer Epidemiol Biomarkers Prev. 2019;28: 1735–1745.

40. Law PJ, Timofeeva M, Fernandez-Rozadilla C, Broderick P, Studd J, Fernandez-Tajes J, et al. Association analyses identify 31 new risk loci for colorectal cancer susceptibility. Nat Commun. 2019;10: 2154.

41. Lawrenson K, Song F, Hazelett DJ, Kar SP, Tyrer J, Phelan CM, et al. Genome-wide association studies identify susceptibility loci for epithelial ovarian cancer in east Asian women. Gynecol Oncol. 2019;153: 343–355.

42. Ulirsch JC, Lareau CA, Bao EL, Ludwig LS, Guo MH, Benner C, et al. Interrogation of human hematopoiesis at single-cell and single-variant resolution. Nat Genet. 2019;51: 683–693.

43. Onuchic V, Lurie E, Carrero I, Pawliczek P, Patel RY, Rozowsky J, et al. Allele-specific epigenome maps reveal sequence-dependent stochastic switching at regulatory loci. Science. 2018;361. doi:10.1126/science.aar3146

44. Studd JB, Yang M, Li Z, Vijayakrishnan J, Lu Y, Yeoh AE-J, et al. Genetic predisposition to B-cell acute lymphoblastic leukemia at 14q11.2 is mediated by a CEBPE promoter polymorphism. Leukemia. 2019;33: 1–14.

45. Gan KA, Carrasco Pro S, Sewell JA, Fuxman Bass JI. Identification of Single Nucleotide Non-coding Driver Mutations in Cancer. Front Genet. 2018;9: 16.

46. Pierce S, Coetzee GA. Parkinson's disease-associated genetic variation is linked to quantitative expression of inflammatory genes. PLoS One. 2017;12: e0175882.

47. Gonsky R, Fleshner P, Deem RL, Biener-Ramanujan E, Li D, Potdar AA, et al. Association of Ribonuclease T2 Gene Polymorphisms With Decreased Expression and Clinical Characteristics of Severity in Crohn's Disease. Gastroenterology. 2017;153: 219–232.

48. Feng Y, Rhie SK, Huo D, Ruiz-Narvaez EA, Haddad SA, Ambrosone CB, et al. Characterizing Genetic Susceptibility to Breast Cancer in Women of African Ancestry. Cancer Epidemiol Biomarkers Prev. 2017;26: 1016–1026.

49. Phelan CM, Kuchenbaecker KB, Tyrer JP, Kar SP, Lawrenson K, Winham SJ, et al. Identification of 12 new susceptibility loci for different histotypes of epithelial ovarian cancer. Nat Genet. 2017;49: 680–691.

50. Law PJ, Berndt SI, Speedy HE, Camp NJ, Sava GP, Skibola CF, et al. Genome-wide association analysis implicates dysregulation of immunity genes in chronic lymphocytic leukaemia. Nat Commun. 2017;8: 14175.

51. Molineros JE, Yang W, Zhou X-J, Sun C, Okada Y, Zhang H, et al. Confirmation of five novel susceptibility loci for systemic lupus erythematosus (SLE) and integrated network analysis of 82 SLE susceptibility loci. Hum Mol Genet. 2017;26: 1205–1216.

52. Zhang T, Xu M, Makowski MM, Lee C, Kovacs M, Fang J, et al. Promoter Mutations Ablate GABP Transcription Factor Binding in Melanoma. Cancer Res. 2017;77: 1649–1661.

53. Rand KA, Song C, Dean E, Serie DJ, Curtin K, Sheng X, et al. A Meta-analysis of Multiple Myeloma Risk Regions in African and European Ancestry Populations Identifies Putatively Functional Loci. Cancer Epidemiol Biomarkers Prev. 2016;25: 1609–1618.

54. Coetzee SG, Pierce S, Brundin P, Brundin L, Hazelett DJ, Coetzee GA. Enrichment of risk SNPs in regulatory regions implicate diverse tissues in Parkinson's disease etiology. Sci Rep. 2016;6: 30509.

55. Zhigulev A, Norberg Z, Cordier J, Spalinskas R, Bassereh H, Björn N, et al. Enhancer mutations modulate the severity of chemotherapy-induced myelosuppression. Life Sci Alliance. 2024;7. doi:10.26508/lsa.202302244

56. Jeong R, Bulyk ML. Blood cell traits' GWAS loci colocalization with variation in PU.1 genomic occupancy prioritizes causal noncoding regulatory variants. Cell Genom. 2023;3: 100327.

57. Thomas SL, Xu T-H, Carpenter BL, Pierce SE, Dickson BM, Liu M, et al. DNA strand asymmetry generated by CpG hemimethylation has opposing effects on CTCF binding. Nucleic Acids Res. 2023;51: 5997–6005.

58. Kosoy R, Fullard JF, Zeng B, Bendl J, Dong P, Rahman S, et al. Genetics of the human microglia regulome refines Alzheimer's disease risk loci. Nat Genet. 2022;54: 1145–1154.

59. Khetan S, Kales S, Kursawe R, Jillette A, Ulirsch JC, Reilly SK, et al. Functional characterization of T2D-associated SNP effects on baseline and ER stress-responsive β cell transcriptional activation. Nat Commun. 2021;12: 5242.

60. Linder RA, Zabanavar B, Majumder A, Hoang HC-S, Delgado VG, Tran R, et al. Adaptation in Outbred Sexual Yeast is Repeatable, Polygenic and Favors Rare Haplotypes. Mol Biol Evol. 2022;39. doi:10.1093/molbev/msac248

61. D'Aurizio R, Catona O, Pitasi M, Li YE, Ren B, Nicolis SK. Bridging between Mouse and Human Enhancer-Promoter Long-Range Interactions in Neural Stem Cells, to Understand Enhancer Function in Neurodevelopmental Disease. Int J Mol Sci. 2022;23. doi:10.3390/ijms23147964

62. Mononen J, Taipale M, Malinen M, Velidendla B, Niskanen E, Levonen A-L, et al. Genetic variation is a key determinant of chromatin accessibility and drives differences in the regulatory landscape of C57BL/6J and 129S1/SvImJ mice. Nucleic Acids Res. 2024;52: 2904–2923.

63. Li J, Xiang Y, Zhang L, Qi X, Zheng Z, Zhou P, et al. Enhancer-promoter interaction maps provide insights into skeletal muscle-related traits in pig genome. BMC Biol. 2022;20: 136.

64. Krishnan J, Seidel CW, Zhang N, Singh NP, VanCampen J, Peuß R, et al. Genome-wide analysis of cis-regulatory changes underlying metabolic adaptation of cavefish. Nat Genet.

2022;54: 684–693.

65. Vespasiani DM, Jacobs GS, Cook LE, Brucato N, Leavesley M, Kinipi C, et al. Denisovan introgression has shaped the immune system of present-day Papuans. PLoS Genet. 2022;18: e1010470.

66. Kuderna LFK, Ulirsch JC, Rashid S, Ameen M, Sundaram L, Hickey G, et al. Identification of constrained sequence elements across 239 primate genomes. Nature. 2024;625: 735–742.

67. Coetzee SG, Coetzee GA, Hazelett DJ. motifbreakR: an R/Bioconductor package for predicting variant effects at transcription factor binding sites. Bioinformatics. 2015;31: 3847–3849.

68. Shannon P RM. MotifDb: An Annotated Collection of Protein-DNA Binding Sequence Motifs. R package version 1.46.0. In: Bioconductor.org [Internet]. 2024. Available: https://www.bioconductor.org/packages/release/bioc/html/MotifDb.html

69. Wingender E, Schoeps T, Haubrock M, Dönitz J. TFClass: a classification of human transcription factors and their rodent orthologs. Nucleic Acids Res. 2015;43: D97–102.

70. Nassar LR, Barber GP, Benet-Pagès A, Casper J, Clawson H, Diekhans M, et al. The UCSC Genome Browser database: 2023 update. Nucleic Acids Res. 2023;51: D1188–D1195.

71. Li D, Hsu S, Purushotham D, Sears RL, Wang T. WashU Epigenome Browser update 2019. Nucleic Acids Res. 2019;47: W158–W165.

72. Li D, Purushotham D, Harrison JK, Hsu S, Zhuo X, Fan C, et al. WashU Epigenome Browser update 2022. Nucleic Acids Res. 2022;50: W774–W781.

73. Robinson JT, Thorvaldsdóttir H, Winckler W, Guttman M, Lander ES, Getz G, et al. Integrative genomics viewer. Nat Biotechnol. 2011;29: 24–26.

74. Thorvaldsdóttir H, Robinson JT, Mesirov JP. Integrative Genomics Viewer (IGV): high-performance genomics data visualization and exploration. Brief Bioinform. 2012;14: 178–192.

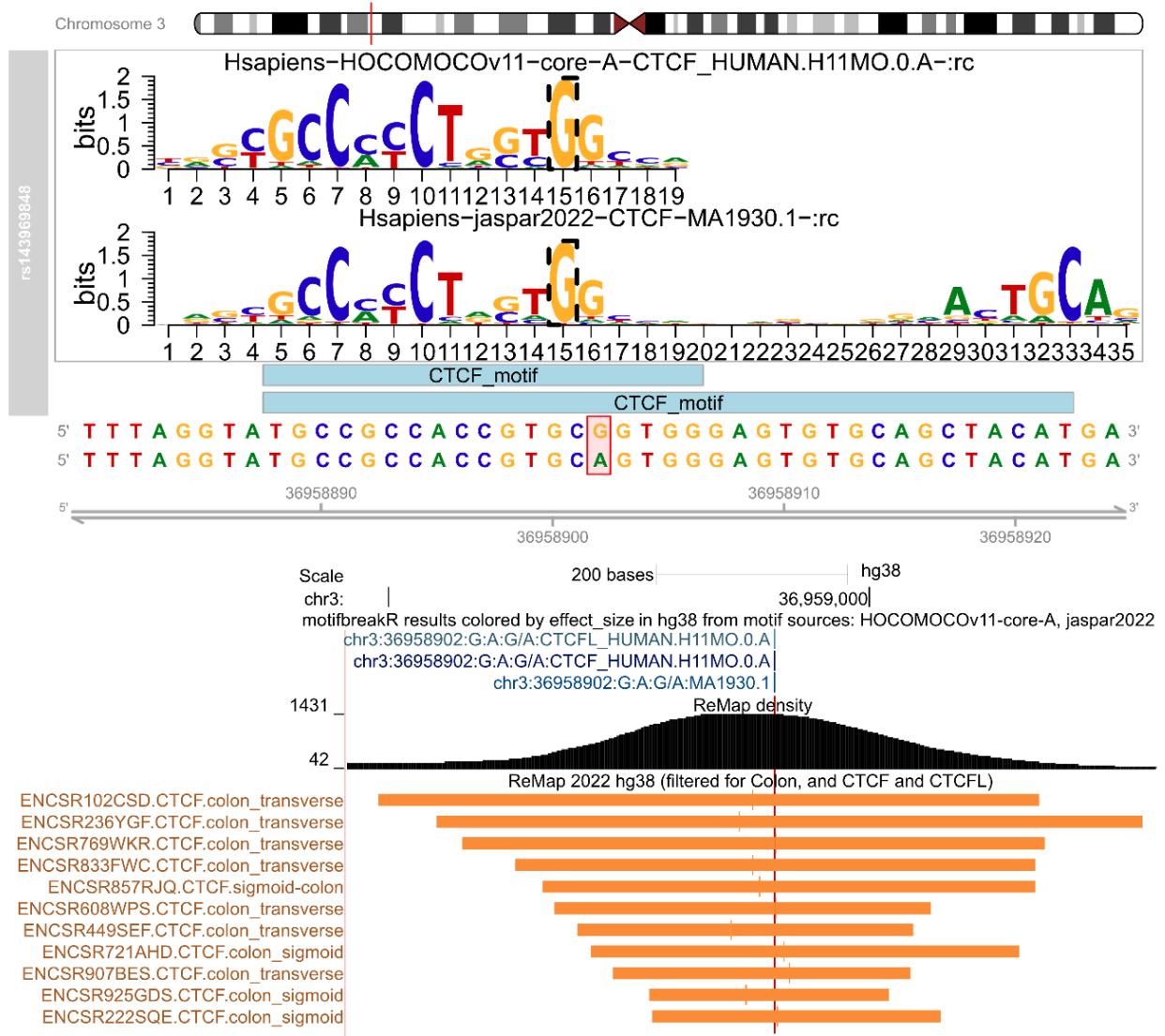

**Figure 1.**

**A.** Here, we represent four distinct ways that an indel can interrupt a motif and the coordinates that describe the position of the motif relative to the indel. A dark gray box highlights the consensus sequence of the motif, and the light grey box indicates the position of the indel.

**B.** A genome browser shot showing the export of *motifbreakR* results colored by the effect size of the variant. A darker color indicates a stronger effect. The remaining tracks illustrate native tracks available on the UCSC genome browser, "ReMap density" (the density of ReMap2022 results), and a representative sample of individual ChIP-Seq binding peaks for CTCF in colon samples from the same browser track. The *motifbreakR* results indicate the potential for a CTCF motif to be disrupted by the alternate allele of the variant, and ReMap2022 indicates evidence that CTCF can be found binding to the site of the variant.